\documentstyle[12pt]{article}
\begin{document}
\pagestyle{empty}
\setlength{\oddsidemargin}{0.5cm}
\setlength{\evensidemargin}{0.5cm}
\setlength{\footskip}{1.5cm}
\renewcommand{\thepage}{-- \arabic{page} --}
\newcommand{\beq}{\begin{equation}}
\newcommand{\eeq}{\end{equation}}
\vspace*{1.25cm}

\renewcommand{\thefootnote}{*)}
\centerline{\large\bf Strategy for the 1995 LEP energy scan.}

\vspace*{1.75cm}

\centerline{\sc M. CONSOLI$^{\: a)}$ and M. PICCOLO$^{\: b)}$}

\vspace*{1.75cm}
\centerline{\sl $a)$ Istituto Nazionale di Fisica Nucleare - Sezione
di Catania}
\centerline{\sl Corso Italia, 57 - I 95129 Catania - ITALY}

\vskip 0.3cm
\centerline{\sl $b)$ Laboratori Nazionali dell' INFN di Frascati}
\centerline{\sl Via E. Fermi, 40 - I-00044 Frascati - ITALY }

\vspace*{3cm}

\centerline{ABSTRACT}

\vspace*{0.4cm}
\baselineskip=20pt plus 0.1pt minus 0.1pt
We propose a  procedure to  measure the Z$^{0}$ line shape
and  take full advantage of the
superb performance of the CERN $e^+e^-$ Collider LEP. A precise determination
of the total cross section at 5 energies is needed for a model-independent
analysis of the data and for a precision test of the QED initial state
radiation from the fully inclusive hadronic channel.

\vfill
\newpage
\pagestyle{plain}
\setcounter{footnote}{0}
\baselineskip=21.0pt plus 0.2pt minus 0.1pt

Motivated by the presentations given by T. Camporesi \cite{campo} and
L. Rolandi
\cite{rola} at the 1995 Italian LEP Workshop held in Genova  last April, we
shall discuss in this Letter a
strategy for the 1995 LEP energy scan. Indeed, there
seems to be the possibility that the 1995 scan be performed at precisely
the same energies explored in the past:  peak and  $\pm 2$ GeV off-peak.
(The 1991 data do not add very much given their lower statistical
significance.) We shall
point out that this decision would not, in our opinion,
exploit to the fullest the experimental potentialities of the last Z$^{0}$
LEP run.

The theoretical foundation for a model-independent analysis of the
Z$^{0}$-line shape, as first proposed by Borrelli et al. in \cite{borrelli},
and
motivated only by the general requirements of unitarity and renormalizability
of an underlying quantum field theory, is based on a description of the
total cross sections for the process $e^+e^- \to f \bar f +...$ \cite{note}
in terms of (at least) 4 free parameters.
Namely one has to introduce the Z$^{0}$ mass $M$, its total decay decay width
$\Gamma$, the peak cross section
$\sigma^{(f)}_{peak}={{12\pi}\over{M^2}}{{\Gamma_e\Gamma_f}\over{\Gamma^2}}$
in the given channel and the parameter $R^{(f)}$ controlling the interference
of the Z$^{0}$-resonance with the non resonating background. In this case, this
minimal parametrization introduced in \cite{borrelli} guarantees that the
analysis of the experimental data is model-independent (to a high degree of
accuracy) so that they can be interpreted consistently within a wide class
of theoretical frameworks beyond the standard model minimal structure
(e.g. super-symmetry, additional Z', technicolor and so on). We stress that,
apart from the potential effects of new physics,
the requirement
of model-independence has a deep theoretical meaning. In fact,
only in this case,
one can quote unambiguously the Z$^{0}$ mass, its total and partial decay rates
 without specifying that these are the values of the parameters within the
assumptions of the minimal standard model (i.e. with three fermion generations,
one Higgs doublet, and so on).

Since the Z$^{0}$-line shape is dominated by the $e^+e^-\to hadrons$ data we
shall
mainly concentrate on the hadronic channel. Within the consistent approach of
\cite{borrelli},  the values of the total cross section (at least)
at 5 energies have to be measured to fully constrain the 4 free parameters
$M, \Gamma, \sigma^{(h)}_{peak}, R^{(h)}$ and check the quality of the fit.
At the present, with only 3 energies explored at high statistics, the hadronic
interference parameter $R^{(h)}$ has been fixed at its standard model
value. With the 1995 scan, we could really perform a model-independent
analysis provided the off-peak data are collected at { \it different}
energies (e.g. at the peak and at $\pm 3$ GeV off-peak). We understand
that the choice of running at $\approx$ 2 GeV off-peak is the result of an
optimization which takes into account both the counting rate and the
analyzing power toward the determination of the Z$^{0}$ width. We
believe however, that, with a small degradation in the  analyzing
power, using different and eventually asymmetric energy settings to compensate
the overall event yield to first order, one could gain  a better
understanding of the resonance parameters as a whole.
We stress that,
independently of the interest in determining $R^{(h)}$ from the precise LEP
data
and comparing its experimental value with the various theoretical predictions,
when $R^{(h)}$ is constrained at its standard model value the actual
error on the Z$^{0}$ mass is underestimated since $M$ and $R^{(h)}$ are
strongly
correlated in the fit \cite{borrelli}. Furthermore,to the present level of
accuracy,
the additional uncertainty induced by the presence of an extra free
parameter in the hadronic peak cross section and in the total width should not
be neglected. Finally, this additional uncertainty from the interference
parameters will propagate through all channels and in the various
observables  and
{\it has to be taken into account} for a meaningful extimate of the
$\epsilon$-parameters \cite{altarelli} accounting for potential new physics
effects beyond the minimal standard model.

Quite independently of the above discussion  a high statistics
determination of the cross sections at 5 energies is needed for a precision
test of the pure QED effects around the Z$^{0}$ resonance. At the present, in
the
fully inclusive hadronic channel, the theoretical predictions for initial state
soft-photon and hard-photon radiation are believed to be under control at the
level of 0.2-0.3$\%$. A check of this statement requires to investigate the
quality of the $\chi^2$ by comparing the theoretical cross section with the
experimental data. Since there are (at least) three free parameters
$M,\Gamma,\sigma^{(h)}_{peak}$
such a test would not be possible if only
three cross section measurements (even with {\it infinite} precision)
at three different energies were
available.

As a final remark we stress that
in the more delicate case of the forward-backward asymmetries, their pole
values at $\sqrt s=M$ are determined from a fit with energy dependent
formulae. It is quite conceivable that by measuring the asymmetries at two new
energies the fit would become considerably more precise and
potential sources of systematic error could be identified.

In conclusion: an alternative  procedure of energy scanning for the
Z$^{0}$ pole  is suggested, to take  full
advantage of the superb performance of the CERN $e^+e^-$ Collider LEP.
A precise
determination of the total cross sections at 5 energies would be extremely
beneficial  for a
model-independent analysis of the data and for a precision test of the QED
initial state radiation from the fully inclusive hadronic channel.



\newpage
\begin{thebibliography}{99}
\bibitem{campo}
T. Camporesi, {\it LEP: future perspectives
for high luminosity and high energy},
Talk given at the 1995 Italian LEP
Workshop, Genova, 10-12 April 1995.
%
\bibitem{rola}
L. Rolandi, {\it Energy calibration at LEP},
Talk given at the 1995 Italian LEP
Workshop, Genova, 10-12 April 1995.
%
\bibitem{borrelli}
A. Borrelli, M. Consoli, L. Maiani and R. Sisto, Nucl. Phys. {\bf B333} (1990)
357.
%
\bibitem{note}
The dots indicate that the final state ($f\neq e$) is electromagnetically
inclusive.
%
\bibitem{altarelli}
G. Altarelli, R. Barbieri and S. Jadach, Nucl. Phys. {\bf B369} (1992) 3;
{\bf B376} (1992) 444(E).
%
\end{thebibliography}
\end{document}